\documentclass[aps,prl,twocolumn,noshowpacs,amsmath,amssymb]{revtex4-1}
\usepackage{graphicx}
\usepackage{amsmath}
\usepackage[usenames]{color}
\usepackage{bm}
\usepackage[normalem]{ulem}

\makeatletter
\newcommand\footnoteref[1]{\protected@xdef\@thefnmark{\ref{#1}}\@footnotemark}
\makeatother

\begin{document}
\title{Stability and roughness of interfaces in mechanically-regulated tissues}
\date{\today}
\author{John.\ J.\ Williamson and Guillaume Salbreux}
\affiliation{The Francis Crick Institute, 1 Midland Road, London NW1 1AT, UK}

\begin{abstract}
Cell division and death can be regulated by the mechanical forces within a tissue. We study the consequences for the stability and roughness of a propagating interface, by analysing a model of mechanically-regulated tissue growth in the regime of small driving forces. For an interface driven by homeostatic pressure imbalance or leader-cell motility, long and intermediate-wavelength instabilities arise, depending respectively on an effective viscosity of cell number change, and on substrate friction. A further mechanism depends on the strength of directed motility forces acting in the bulk. 
We analyse the fluctuations of a stable interface subjected to cell-level stochasticity, and find that mechanical feedback can help preserve reproducibility at the tissue scale. Our results elucidate mechanisms that could be important for orderly interface motion in developing tissues.

\end{abstract}

\maketitle

Interfaces are ubiquitous in tissue biology, between a tissue and its environment \cite{Hong2016, Ravasio2015, Basan2013} or between cell populations \cite{Baker2011, Ninov2007, Vincent2013, Gogna2015, Podewitz2016}. There is great interest in how interfaces propagate smoothly or maintain their shape in the face of cell proliferation and renewal \cite{Hong2016, Marianes2013, Ninov2007, Landsberg2009, MC2017, Curtius2017}, for example by line tension acting at tissue boundaries \cite{Sussman2018, Cayuso2015, Landsberg2009, Bielmeier2016}. 

Theoretical efforts have focused on contour instabilities in cancer \cite{Tracqui2009, Cristini2005, Poplawski2010, Ciarletta2011, Benamar2011}, branching \cite{Lubkin1995, Miura2015} or folding \cite{Bayly2013}, and wound healing \cite{Zimmermann2014, Tarle2015, Basan2013}. In models that include nutrient diffusion, protruding regions access more nutrient, triggering further growth \cite{Castro2005, Poplawski2010, Cristini2005}, reminiscent of the Mullins-Sekerka instability in non-living systems \cite{Mullins1963}. An epithelium-stroma interface could form undulations due to mechanical stresses from cell turnover \cite{Basan2011, Risler2013}, while a Saffman-Taylor-like instability based on viscosity contrast has been proposed to underlie branching in the developing lung \cite{Lubkin1995}. A recent cell-based simulation of imbalanced mechanically-regulated growth between two epithelia observed a stable interface, and quantified its roughness \cite{Podewitz2016}. A related simulation of cells in an inert medium found finger-like protrusions, arising for higher friction in the medium relative to the cells \cite{Drasdo2012, Lorenzi2017}. Ref.\ \cite{Risler2015} calculated the steady-state surface fluctuations of a non-growing tissue maintained in its homeostatic state. 

In tissue replacement, such as in the developing \textit{Drosophila} abdominal epidermis \cite{Ninov2010, Bischoff2012}, interface propagation occurs. This may be driven by imbalances in pressure associated with cell division, and/or directed cell motility, which cause the expansion of one tissue at the other's expense.

In this Letter, we ask whether factors that drive an interface's propagation can also affect its stability and roughness. We are particularly interested in the consequences of mechanically-regulated cell division and death for the behaviour of interfaces. If cell number change is sensitive to mechanical forces \cite{Montel2011, Streichan2014, Puliafito2012, Benham2015, Aegerter2007, Pan2016, Levayer2016, Gudipaty2017, MC2017, Fletcher2018, Legoff2016} it leads to a ``homeostatic pressure'' \cite{Basan2009, Ranft2014, Recho2016, Podewitz2016} which can drive interface propagation without coherently-directed cell motility forces. Alternatively, active, directed migration is proposed to drive interface motion in wound healing \cite{Cochet2014, Ravasio2015, Poujade2007} or tissue replacement \cite{Bischoff2012}. 
\begin{figure}
\begin{center}
\includegraphics[width=8.6cm]{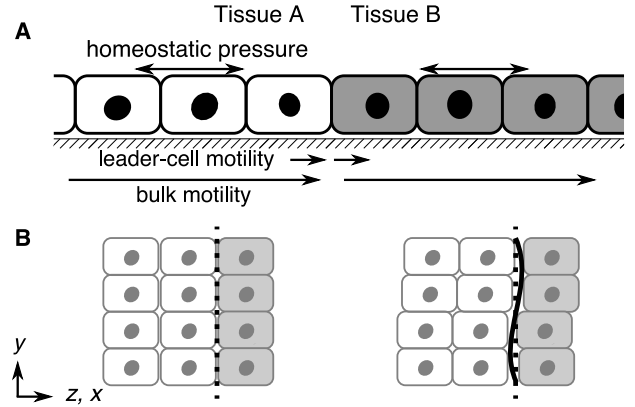}
\caption{\label{fig:schematic} A)\ Side-on schematic of competing epithelial tissues. Each tissue is described with coarse-grained fields of cell density $\rho$, stress $\sigma_{ij}$ and velocity $v_i$. Each has an elastic modulus $\chi$ (Eq.\ \ref{eqn:elasticstress}), substrate friction $\xi$ (Eq.\ \ref{eqn:forcebalance}), and division/death responsive to strain on a timescale $\tau$ (Eq.\ \ref{eqn:divrate}).
B)\ Top-down view of the tissues illustrating an interface contour fluctuation. We use a 2D description, with $z$ a coordinate parallel to $x$ in the comoving frame of the interface.}
\end{center}
\end{figure}
We study a model of competing epithelial tissues, with an interface driven by homeostatic pressure or by directed motility, acting either at the interface (a ``leader-cell'' limit) or in bulk \footnote{See Supplemental Material for further theoretical calculations, estimates of parameters, and Refs.\ \cite{Saffman1958,Edwards1982,Blanch2017,Bonnet2012,Bambardekar2015,Sepulveda2013,Curtain2009,Traechtler2016,Ansgar}.} \cite{Farooqui2004, Trepat2009, Scarpa2016}. Our results encompass also the cases of stationary interfaces maintained under constant cell renewal \cite{Marianes2013}, and of a single growing tissue \cite{Hong2016}.

We find a Saffman-Taylor-like instability involving substrate friction, and a long-wavelength instability dependent on an effective viscosity of cell number change.
Bulk motile forces induce an instability depending on their strength and direction in each tissue. The free boundary of a growing \textit{single} tissue is generally stabilised by the mechanisms studied here. 
Adding a driving noise to represent, \textit{e.g.}, stochastic cell division, we calculate the roughness and centre-of-mass diffusion of a stably-propagating interface, and find that mechanical feedback can help preserve reproducibility at the tissue scale \cite{Hong2016}. 

We use a 2D hydrodynamic description in terms of cell density and velocity fields. Tissues $A$ and $B$ cover an infinite domain, meeting at a flat interface (Fig.\ \ref{fig:schematic}A). Since we assume a sharp interface, we state equations for a general tissue, unless decorated with $A$ or $B$.

We begin with continuity of the areal cell density, $\rho$,
\begin{equation}
\partial_t \rho + \partial_i (\rho v_i) = k_d \rho ~,
\label{eqn:number}
\end{equation}
\noindent where $v_i$ is the velocity field and
\begin{equation} \label{eqn:divrate}
k_d = \frac{1}{\tau} \frac{\rho_d - \rho}{\rho_d}
\end{equation}
\noindent is an expansion of the net division/death rate about the homeostatic density $\rho_d$, with $\tau$ a characteristic timescale \cite{Basan2009}. The $\rho$ units of each tissue are independent, so we set $\rho_d \equiv 1$.
\noindent We consider a linearised, isotropic elastic stress,
\begin{equation} \label{eqn:elasticstress}
\sigma_{ij} =\sigma \delta_{ij}\, , \, \sigma \equiv \sigma_h - \chi \Delta \rho~,
\end{equation}
\noindent where $\sigma_h$ is a tissue's homeostatic stress, $\chi$ its elastic modulus and $\Delta \rho \equiv \rho - 1$. The homeostatic pressure imbalance is $-\Delta \sigma_h \equiv -(\sigma_{hA} - \sigma_{hB})$.
The quantity $\chi \tau$ is an \textit{effective bulk viscosity} for cell number change \cite{Lubkin1995};\ on a timescale $\tau$, a tissue loses its elastic character as cells are lost or created in response to elastic stress \cite{Ranft2010, Risler2015}. Based on parameter estimates (see supplement \cite{Note1}), we neglect viscous stresses $\sim \bar{\eta} \partial_i v_j$, anticipating $\chi \tau \gg \bar{\eta}$ \cite{Recho2016,Note1} . 
Force balance expresses the tissue velocity as
\begin{equation} \label{eqn:forcebalance}
v_i = \xi^{-1} (\partial_i \sigma + f_i)~,
\end{equation}
for substrate friction $\xi$ and a density of active motility forces $f_i = \delta_{ix} f$ directed normal to the interface. ``Leader-cell'' motility \textit{at} the interface gives an effective contribution to $\Delta \sigma_h$ \cite{Note1}. We thus take $f$ in Eq.\ \ref{eqn:forcebalance} as uniform in a given tissue, to account for ``bulk'' directed motility forces, as may arise from cryptic lamellipodia away from tissue edges \cite{Farooqui2004, Trepat2009, Scarpa2016}.

\begin{figure}
\begin{center}
\includegraphics[width=8.6cm]{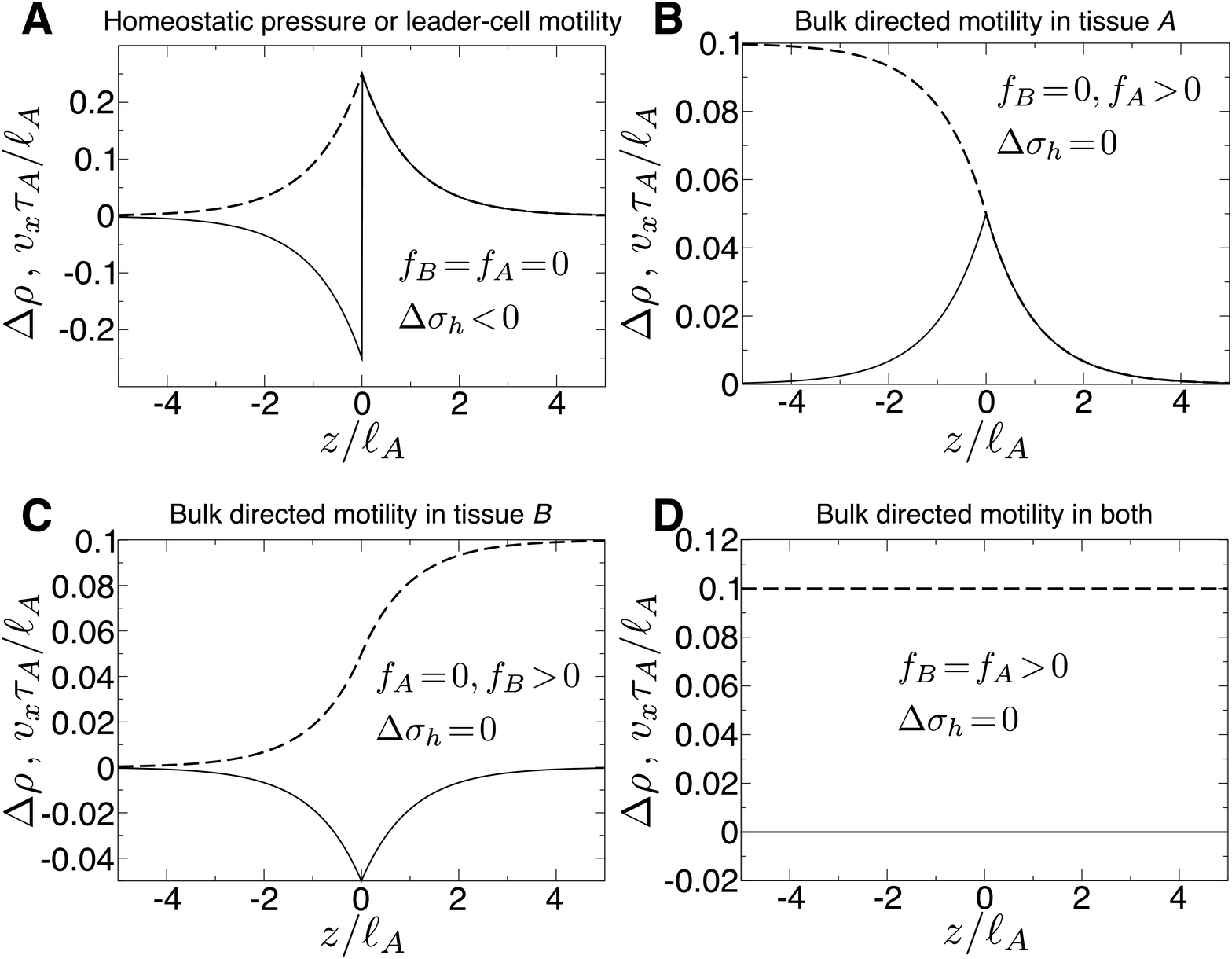}
\caption{\label{fig:rhonvel} A) Steady state density perturbation profile $\Delta \rho$ (solid line) and velocity $v_x$ (dashed) for tissues $A$ ($z < 0$) and $B$ ($z > 0$).
Parameters:\ homeostatic pressure difference $-\Delta \sigma_h = 0.5\chi_A$, frictions $\xi_B = \xi_A$, elastic moduli $\chi_B = \chi_A$, division/death timescales $\tau_B = \tau_A$, bulk motility forces $f_A = f_B = 0$. B) As A, but with $\Delta \sigma_h = 0$, $f_A = 0.1 \chi_A / \ell_A$. C) As A, but with $\Delta \sigma_h = 0$, $f_B = 0.1 \chi_A / \ell_A$. D) As A, but with $\Delta \sigma_h = 0$, $f_A = f_B = 0.1 \chi_A / \ell_A$. In this particular case the tissue moves uniformly and the density perturbation cancels to zero.}
\end{center}
\end{figure}

\paragraph{Moving steady state.}
We first solve for the steady propagation of a flat interface (cf.\ Refs.\ \cite{Recho2016, Podewitz2016}). The comoving coordinate is $z \equiv x - V_0 t$, with $V_0$ the velocity of the interface at $z_0=0$, propagating in $z$. 
Assuming driving forces small enough that nonlinear terms in $\Delta \rho$, $v$ can be neglected in Eq. \ref{eqn:number}, we write
\begin{equation}
\partial_t \Delta \rho + \partial_z v_z + \partial_y v_y = -\frac{1}{\tau} \Delta \rho~. \label{eqn:startingpoint}
\end{equation}
where $v_z\equiv v_x-V_0$.
The resulting propagating steady state is derived in the supplement \cite{Note1} by matching the tissues' stress and velocity at the interface. Density perturbations $\propto e^{\pm z / \ell}$ decay from the interface (Fig.\ \ref{fig:rhonvel}) governed by each tissue's hydrodynamic length $\ell = \sqrt{\chi \tau /\xi}$.
Their sign (see Eq.\ S2 \cite{Note1}) depends on $-\Delta \sigma_h$, and on $\Delta v_f \equiv f_A /\xi_A  - f_B /\xi_B$, a difference in ``bare'' velocities $f / \xi$ associated to the bulk directed motilities. 
In Fig.\ \ref{fig:rhonvel}A, the growing tissue has decreased density so, by Eq.\ \ref{eqn:divrate}, is proliferative near the interface, while the shrinking tissue has increased density so undergoes net apoptosis near the interface. In Fig.\ \ref{fig:rhonvel}B, \textit{both} tissues are apoptotic near the interface. The steady interface velocity,
\begin{equation} \label{eqn:steadyvel}
V_0 \equiv v_{x} \rvert_{z=0} =  \frac{-\Delta \sigma_h + \ell_A f_A + \ell_B f_B}{\xi_A \ell_A + \xi_B \ell_B}~,
\end{equation}
\noindent 
is, for $f_A\!=\!f_B=0$ and $\xi_B\!=\!\xi_A$, that found in Ref.\ \cite{Podewitz2016}. To justify ignoring nonlinear terms, we require that stresses from homeostatic pressure, motility and interface line tension $\gamma$ are small relative to the tissues' elastic moduli:\ $\lvert \Delta\sigma_h \rvert \ll \chi$, $\lvert f/\xi \rvert \ll \ell/\tau$, $\gamma/\ell \ll \chi$.
Much stronger stresses would lead to nonlinear responses and, eventually, tissue rupture \cite{Recho2016, Harris2012, Note1}.

\begin{figure}
\begin{center}
\includegraphics[width=8.6cm]{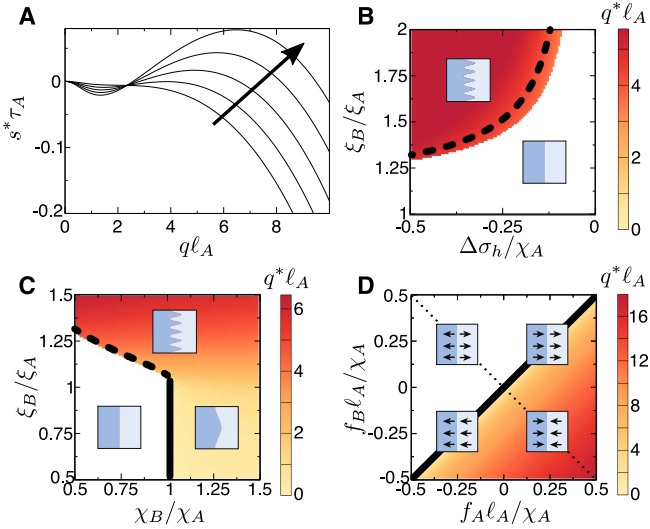}
\caption{\label{fig:approximations}
A) Example numerically-determined dispersion relations. Parameters:\ frictions $\xi_B = 1.5 \xi_A$, moduli $\chi_B = 0.5 \chi_A$, division/death timescales $\tau_B = \tau_A$, bulk motility forces $f_A = f_B = 0$, line tension $\gamma = 0.001 \ell_A \chi_A$. The homeostatic pressure imbalance/leader-cell motility parameter varies, $\Delta \sigma_h = -0.1 \chi_A, -0.2 \chi_A, \dots, -0.5\chi_A$, in the direction of the arrow, crossing a ``type I$_{\rm s}$'' instability transition \cite{Cross1993}.
B) Phase diagram in $(\Delta \sigma_h, \xi_B)$ of the most unstable wavenumber $q^*$ (white if no instability), using $\chi_B = 0.5 \chi_A$, $\tau_B = \tau_A$, $f_A = f_B = 0$, $\gamma = 0.001 \ell_A \chi_A$. The dashed line indicates the approximation of the type I$_{\rm s}$ transition line by Eq.\ \ref{eqn:frictioncriterion}. 
C) Phase diagram in $(\chi_B, \xi_B)$ using $\Delta \sigma_h = -0.5 \chi_A$, $\tau_B = \tau_A$, $f_A = f_B = 0$, $\gamma = 0.001 \ell_A \chi_A$. The meaning of the dashed line is as in C, whereas the solid line indicates the type II$_{\rm s}$ transition (Eq.\ \ref{eqn:visccriterion}).  
D) Phase diagram in bulk directed motilities $(f_A, f_B)$, with other parameters $\xi_B = \xi_A$, $\chi_B = \chi_A$, $\tau_B = \tau_A$, $\gamma = 0.001 \ell_A \chi_A$. The black line is the transition approximated by Eq.\ \ref{eqn:delvfcriterion}. The dotted line is $V_0 = 0$, with the upper half-space $V_0 >0$ (tissue $A$ growing) and the lower $V_0<0$. In each quadrant, cartoons illustrate the direction of the bulk motilities.
}
\end{center}
\end{figure}

\paragraph{Interface stability.}
In the supplement \cite{Note1}, starting from Fourier and Laplace transforms $y\to q$ and $t \to s$ of Eq.\ \ref{eqn:startingpoint}, we perturb the propagating steady state calculated above, to find the fate of an interface fluctuation (Fig.\ \ref{fig:schematic}B) $\delta z_0 = \epsilon(t) \cos (qy)$, where $\epsilon(0)=\epsilon_0$.
For line tension $\gamma \geq 0$ (\textit{e.g.}, increased myosin at heterotypic junctions \cite{Landsberg2009} or a supracellular actin cable \cite{Hayes2017}),
\begin{equation} \label{eqn:smatch}
(\sigma + \delta \sigma) \rvert_{\delta z_0, A} - (\sigma + \delta \sigma) \rvert_{\delta z_0, B}= -\gamma q^2 \delta z_0~,
\end{equation}
\noindent where $\delta  \sigma$ is the deviation from stress $\sigma$ of the propagating steady state. The dominant growth rate is denoted $s^*(q)$, with $s^*<0$, $>0$ indicating stability or instability (in the applicable parameter regime we do not find complex poles, so treat $s^*$ as real \cite{Note1}). Dispersion relations (\textit{e.g.}, Fig.\ \ref{fig:approximations}A) maximised over $q$ yield phase diagrams (Fig.\ \ref{fig:approximations}B,C,D) of the most-unstable wavenumber $q^*$. We approximate the dispersion relation in limits of $q$ \cite{Note1} to find the analytic criteria discussed below.

\paragraph{Interface driven by homeostatic pressure or leader-cell motility.} We first discuss growth of tissue $A$ driven by $\Delta\sigma_h < 0$ (Fig.\ \ref{fig:approximations}A,B,C), without bulk directed motility ($f_A=f_B=0$). Analytic dispersion relations \cite{Note1} show that, for strong enough $\Delta\sigma_h$, the interface is unstable if tissue $B$ has greater friction $\xi_B > \xi_A$ or effective viscosity $\chi_B \tau_B > \chi_A \tau_A$. 
Instability criteria (given $\Delta \sigma_h <0$) are
\begin{align} \label{eqn:frictioncriterion}
- \Delta \sigma_h&\gtrsim \frac{27}{4} \gamma \frac{(\xi_A \ell_B - \xi_B \ell_A)^2 ( \xi_A \ell_A+  \xi_B\ell_B)}{
\ell_A^2 \ell_B ^2 (\xi_B - \xi_A)^3} ~, \notag \\
\xi_B &> \xi_A ~,
\end{align}
and 
\begin{equation} \label{eqn:visccriterion}
- \Delta \sigma_h> \frac{2 \gamma (\xi_A \ell_A+\xi_B \ell_B)}{\chi_B \tau_B -\chi_A \tau_A}~,~\chi_B\tau_B>\chi_A \tau_A~.
\end{equation}
where Eq.\ \ref{eqn:frictioncriterion} is approximate \cite{Note1}.
Two types of transition arise. Fig.\ \ref{fig:approximations}A and Eq.\ \ref{eqn:frictioncriterion} show a ``type I$_{\rm s}$'' transition in the Cross-Hohenberg classification \cite{Cross1993}, where an intermediate band $q_\textrm{min} < q < q_\textrm{max}\, , \, q_\textrm{min} \neq 0$ becomes unstable (``s'' indicates that the instabilities found are stationary, not oscillatory). Fig.\ S2A \cite{Note1} and Eq.\ \ref{eqn:visccriterion} show a ``type II$_{\rm s}$'' transition, with unstable band $0 < q< q_\textrm{max}$ and onset at $q \to 0$.
Then, one expects near threshold that the characteristic wavelength scales with system size. Eqs.\ \ref{eqn:frictioncriterion}, \ref{eqn:visccriterion} are combined with phase diagrams of $q^*$ in Fig.\ \ref{fig:approximations}B,C.

\paragraph{Interface driven by bulk directed motility.} 

We now consider the case $\Delta \sigma_h=0$, with bulk directed motility forces $f_A\neq 0, f_B\neq0$. Instability occurs when $\Delta v_f \equiv f_A /\xi_A  - f_B /\xi_B$ \cite{Note1} satisfies
\begin{equation} \label{eqn:delvfcriterion}
 \Delta v_f>\frac{2\gamma (\ell_A \xi_A+\ell_B\xi_B)}{  \xi_A\xi_B \ell_A \ell _B(\ell_A+\ell_B) }~.
\end{equation}
Fig.\ S2B \cite{Note1} shows dispersion relations crossing the type II$_{\rm s}$ transition of Eq.\ \ref{eqn:delvfcriterion}. The phase diagram in Fig.\ \ref{fig:approximations}D shows that a static interface ($V_0 = 0$), marginally stable for $f=0$ \cite{Note1}, can be stable or unstable depending on the direction of $f_A$, $f_B$.

\paragraph{Single tissue.} The free boundary of a growing single tissue (\textit{e.g.}, epithelium invading empty substrate \cite{Ravasio2015}), is stabilised by the mechanisms studied here (Eqs.\ S20--S22 \cite{Note1}). 
Protrusion formation is often observed in wound-healing, via a number of proposed mechanisms we have not included \cite{Zimmermann2014, Tarle2015, Basan2013, Mark2010}.

\begin{figure}
\begin{center}
\includegraphics[width=8.6cm]{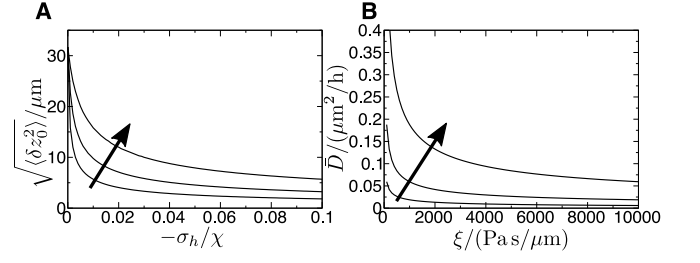}
\caption{\label{fig:roughness}
A) Root-mean-squared interface deviation for a single tissue using example physical parameters (Eq.\ \ref{eqn:roughness1tissue}). The homeostatic pressure/leader-cell motility parameter $\sigma_h$ is varied, and the arrow shows increasing friction $\xi = 10^3, 10^4, 10^5 \, \textrm{Pa} \, \textrm{s}  / \mu \textrm{m}$. Other parameters:\ $\chi = 10^4 \,\textrm{Pa} \, \mu \textrm{m}$, $\tau = 3.6\times10^4\, \textrm{s}$, $\gamma = 10^3\, \textrm{pN}$, 
$L_\perp = 1000\, \mu \textrm{m}$, $D = 0.4\, \mu \textrm{m}^2 / \textrm{h}$.
B) Interface centre-of-mass diffusion coefficient for a single tissue (Eq.\ \ref{eqn:Dsingle}) using example physical parameters. The friction $\xi$ is varied, and the arrow shows increasing elastic modulus $\chi = 10^3, 10^4, 10^5 \, \textrm{Pa}\, \mu \textrm{m}$.
Other parameters:\ $\sigma_h = 0$, $\tau = 3.6\times10^4\, \textrm{s}$, $\gamma = 10^3\, \textrm{pN}$, 
$L_\perp = 1000\, \mu \textrm{m}$, $D = 0.4\, \mu \textrm{m}^2 / \textrm{h}$.
}
\end{center}
\end{figure}

\paragraph{A stable interface subject to noise.} 
Interface propagation and maintenance takes place in the presence of stochasticity in cell divisions, motilities, material parameters, etc. In the supplement \cite{Note1} we model this with i)\ a driving noise in Eq.\ \ref{eqn:startingpoint}, $\partial_t \Delta \rho + \partial_z v_z + \partial_y v_y = -(1/\tau) \Delta \rho + k$ where $\langle k(z,y,t) k(z',y',t') \rangle = D_{A(B)} \delta(z-z') \delta(y-y') \delta(t-t')$, corresponding to a contribution of random cell division; or (ii) a noisy motile force contribution to Eq. \ref{eqn:forcebalance}, $\langle f_i(z,y,t) f_j(z,y,t) \rangle=D^f\delta(y-y')\delta(z-z')\delta(t-t')\delta_{ij}$. Noisy motility could arise, \textit{e.g.}, from `swirling' patterns \cite{Basan2013}, provided that the correlation length of these patterns is small compared to other length scales discussed. We focus here on cell division noise, but find qualitatively similar results for noise on the motile force \cite{Note1}.

Excluding the $q=0$ mode (discussed below), we calculate the correlation function $\langle \delta z_0 (y+y', t+t') \delta z_0 (y,t) \rangle$ of an interface in the stable parameter regime. The saturation (late-time) roughness as the system size in $y$, $L_{\perp}$, becomes large, is
\begin{align} \label{eqn:roughness2tissues}
&\langle \delta z_0(y,t)^2 \rangle \sim \frac{L_{\perp} (\xi_A^2 \ell_A^3 D_A + \xi_B^2 \ell_B^3 D_B)}{ 2 \pi^2\mathcal{N}}~, \notag \\
 &\mathcal{N} \equiv\, 2(\xi_A \ell_A + \xi_B \ell_B ) \gamma - (\chi_A \tau_A - \chi_B \tau_B) \Delta \sigma_h \notag \\ 
 &- \xi_A \xi_B \ell_A \ell_B (\ell_A + \ell_B) \Delta v_f~.
\end{align}
The dependence on $L_{\perp}$ is as in 1-dimensional Edwards-Wilkinson deposition \cite{Antal1996}. The positive denominator $\mathcal{N}$ is expanded in the applicable regime of small $\Delta v_f, \gamma, \Delta \sigma_h$.
Roughness can be reduced by three now-familiar mechanisms:\ line tension $\gamma >0$;\ stabilising effective viscosity contrast $\chi_A \tau_A > \chi_B \tau_B$, $\Delta \sigma_h <0$;\ stabilising bulk motilities $\Delta v_f <0$. 
For identical tissues without line tension, the ``interface'' is an arbitrary line in the tissue:\ Eq.\ \ref{eqn:roughness2tissues} then diverges, \textit{i.e.}, the roughness grows indefinitely. Identical tissues with line tension yield $\langle \delta z_0^2\rangle\sim L_{\perp} \chi \tau D/(4\pi^2 \gamma)$, so that mechanical regulation (\textit{i.e.}, smaller $\tau$) reduces boundary roughness. This is true also for a single tissue, 
\begin{equation} \label{eqn:roughness1tissue}
\langle \delta z_0^2 \rangle \sim L_{\perp} \xi^2 \ell^3 D / \left(2\pi^2 (2 \xi \ell \gamma - \chi \tau \sigma_h)\right)~,
\end{equation}
where if the tissue is growing ($\sigma_h <0$) the roughness is decreased. Fig.\ \ref{fig:roughness}A shows this behaviour quantitatively for estimated physical values of the parameters \cite{Note1}.

The $q=0$ mode leads to an effective diffusion coefficient for the interface centre-of-mass,
\begin{align} \label{eqn:D}
\bar{D} \sim &\frac{1}{L_{\perp}}   \frac{\xi_B^2 \ell_B^3 D_B +  \xi_A^2 \ell_A^3 D_A}{ 4 (\xi_A \ell_A + \xi_B \ell_B)^2}  ~,
\end{align}
and for a single tissue,
\begin{align} \label{eqn:Dsingle}
\bar{D} \sim \frac{\ell D}{ 4L_{\perp}}  ~.
\end{align}
The behaviour of Eq.\ \ref{eqn:Dsingle} for varying friction $\xi$ and elastic modulus $\chi$ is shown in Fig.\ \ref{fig:roughness}B.
These equations control the accumulating uncertainty in tissue size as the interface centre-of-mass progressively diffuses away from its noise-free trajectory. A larger friction coefficient leads to more precise growth (Fig.\ \ref{fig:roughness}B) but decreases the velocity (Eq.\ \ref{eqn:steadyvel}), which suggests trade-offs might be necessary to optimise the speed and precision of growth.

\paragraph{Discussion.} \label{sec:discussion}
Given experimental evidence of mechanically-regulated cell number change \cite{Montel2011, Streichan2014,Puliafito2012, Benham2015, Aegerter2007, Pan2016, Levayer2016, Gudipaty2017, MC2017, Fletcher2018}, models of the type used here are widely studied \cite{Ranft2014, Basan2009, Recho2016, Podewitz2016, Lorenzi2017}. 
There is much interest in mechanisms of boundary maintenance between cell populations \cite{Landsberg2009, Hayes2017, Taylor2017}.
Recent simulations showed how the topology of cellular interactions can stabilise anomalously smooth interfaces \cite{Sussman2018}, while experiments suggest that interface maintenance is not only local but is connected to mechanical waves and jamming processes deep within neighbouring tissues \cite{Rodriguez2017}. 
Our results add to this picture, showing that mechanically-regulated cell number change within in the tissue bulk can exert an important influence on the properties of interfaces. {We have shown how the forces driving overall interface propagation can also generate instabilities, and affect the response of interfaces to cell-level stochasticity.} 

A Saffman-Taylor-like instability based on substrate friction \cite{Drasdo2012, Lorenzi2017} (Eqs.\ \ref{eqn:frictioncriterion}, S17 \cite{Note1}) accords with the tumour literature, where tissues with weaker cell-matrix adhesions tend to be more invasive \cite{Frieboes2006}.
A longer-wavelength instability (Eq.\ \ref{eqn:visccriterion}) occurs if the effective bulk viscosity for cell number change is smallest in the growing tissue. 
Cell-based simulations \cite{Podewitz2016} could explore our predictions, which could in turn be extended to include, \textit{e.g.}, cell growth anisotropy, proposed to play a role in the stable interfaces found in Ref.\ \cite{Podewitz2016}.

The effects of bulk directed motility depend on $\Delta v_f$ (Fig.\ \ref{fig:approximations}D). 
Repulsive migration, known to occur due to Eph and ephrin signalling \cite{Cayuso2015, Taylor2017}, should yield $\Delta v_f < 0$, favouring stability.
In \textit{Drosophila} abdominal epidermis, larval epithelial cells being replaced by histoblasts are proposed to actively migrate away from the propagating interface \cite{Bischoff2012}. This motility force would promote stability, presumably desirable to ensure reproducible, well-controlled tissue replacement. This \textit{Drosophila} system, or model experiments \cite{Rodriguez2017}, could be used to test our theory by perturbing, \textit{e.g.}, motility, substrate friction, or cell division, and observing the effect on interfaces.

We found that mechanical feedback can help to smooth a stable interface in the presence of noise, as well as determining how quickly the interface centre-of-mass diffuses away from its noise-free position. 
These findings are relevant to the question of how tissue-level reproducibility is achieved despite cell-level stochasticity \cite{Hong2016}.

\acknowledgments

We thank Anna Ainslie, John Robert Davis, Federica Mangione and Nic Tapon for discussions. 
J.~J.~W.\ acknowledges support by a Wellcome Trust Investigator award to Dr
Nic Tapon (107885/Z/15/Z), and acknowledges discussions with Ruth Curtain, Claire McIlroy, Pasha Tabatabai and Ansgar Tr\"{a}chtler.
G.~S.\ and J.~J.~W. acknowledge support by the Francis Crick Institute which receives its core funding from Cancer Research UK (FC001317, FC001175), the UK Medical Research Council (FC001317, FC001175), and the Wellcome Trust (FC001317, FC001175).

\bibliography{bibliography}
\end{document}